\documentclass[12pt]{article}
\usepackage{graphicx}
\begin{document}
\rightline{NKU-2014-SF2}
\bigskip

\newcommand{\be}{\begin{equation}}
\newcommand{\ee}{\end{equation}}
\newcommand{\noi}{\noindent}
\newcommand{\refb}[1]{(\ref{#1})}
\newcommand{\ra}{\rightarrow}
\newcommand{\bib}{\bibitem}

\begin{center}
{\Large\bf  Black holes in massive  gravity: Quasi-normal modes of scalar perturbations}

\end{center}
\hspace{0.4cm}
\begin{center}
Sharmanthie Fernando \footnote{fernando@nku.edu} and Tyler Clark \footnote{clarkt3@nku.edu}\\
{\small\it Department of Physics \& Geology}\\
{\small\it Northern Kentucky University}\\
{\small\it Highland Heights}\\
{\small\it Kentucky 41099}\\
{\small\it U.S.A.}\\

\end{center}

\begin{center}
{\bf Abstract}
\end{center}

We have studied  quasinormal modes of  scalar perturbations of a black hole in massive gravity.   The parameters of the theory, such as the mass of the black hole, the scalar charge of the black hole and the spherical harmonic index  is varied to see how the corresponding quasinormal frequencies change. We have also studied the massive scalar field perturbations. Most of the work is done using WKB approach while  sections are devoted to compute quasinormal modes via the unstable null geodesics approach and the P$\ddot{o}$schl-Teller approximation. Comparisons are done with the Schwarzschild black hole.

\hspace{0.7cm} 


\section{ Introduction}

Perturbation of black hole geometry takes a prominent role in research in gravitational physics. The founders of black hole perturbation were Regge and Wheeler: they studied the metric perturbation of Schwarzschild black hole  and showed that the radial part of the equation for axial perturbation  is similar to the Schr$\ddot{o}$dinger equation in 1957 \cite{regge}. Since then there have been  hundreds of papers which have focused on obtaining such equations for variety of black holes and fields of various spins.    A nice review on advanced methods in black hole perturbations is in \cite{pani2}.

When a black hole is subjected to non-radial perturbation, at the intermediate stage, it oscillates with complex frequencies. Such oscillations are  called quasi-normal modes (QNM). These modes are characterized by a set of discrete (usually incomplete) complex frequencies which depends on the properties of the black hole such as mass, spin and the charge.

The studies of these frequencies have drawn intense attention due to variety of reasons.  One of the prominent reasons is the ongoing attempts to observe gravitational waves. If gravitational wave detectors such as LIGO, VIRGO and LISA will be able to detect gravitational wave frequencies emitted by an oscillating black hole, the mass and the angular momentum of the black hole will be able to be determined. An interesting review of gravitational wave astronomy is given by Ferrari and Gualtrieri \cite{ferra}. Also, quasinormal modes of black holes which are asymptotically anti-de Sitter are studied in the context of AdS/CFT correspondence \cite{alex}. The electromagnetic and gravitational quasinormal frequencies of anti-de Sitter black holes are related to the poles of retarted correlation functions in the dual conformal field theory. Another reason to study quasinormal modes was inspired by an interesting conjecture by Hod \cite{hod}. According to the conjecture, the asymptotic quasinormal modes can shed information on quantum properties of the black hole. Two works related to this conjecture are \cite{bronco} and \cite{taka}. Two excellent reviews on quasinormal modes are given  by Konoplya and Zhidenko  \cite{kono1} and Berti et.al. \cite{cardoso1}.

It is impossible to make references to all the work  related to QNM in the literature. Instead, we will refer to some recent interesting work. Quasinormal modes of the Myers-Perry black hole were studied in \cite{oscar}. Quasinormal modes of a naked singularity in the Reissner-Nordstrom space-time were studied in \cite{chirenti}. Quasinormal modes are not only a subject for black holes. Boson stars, which are considered as candidates for dark mater does emit QNM as studied by Macedo et.al \cite{pani}. The electromagnetic quasinormal modes of a rotating black string and the AdS/CFT correspondence were studied in \cite{alex}. Studies of quasinormal modes of regular black holes which does not have singularity at the origin were studied by \cite{fernando1} \cite{hong} \cite{lemos}. Quasinormal modes in de Sitter space have been studied by Tanhayi \cite{reza}. Studies of QNM's in acoustic black holes were presented by Lemos \cite{lemo}.

Massive gravity has attracted great interest in the recent past. Various models where the graviton, which is a spin 2 field, acquire a mass have been considered. The first attempt to give a mass to the graviton was done by  Fierz and Pauli \cite{pauli} in 1939. An excellent review on various theories of massive gravity is given  in \cite{ claudia}.

Among all massive gravity theories, one of the attractive theories are the Lorentz-violating massive gravity theory. In this model, Lorentz symmetry is spontaneously broken by four scalar fields minimally  coupled to gravity through derivative couplings. Theses scalar fields are called Goldstone fields. When the Lorentz symmetry is broken, the graviton acquire a mass. This is very similar to the Higgs mechanism. An excellent review of Lorentz violating theories of massive gravity is found in \cite{dubo} \cite{ruba2}

The theory of massive gravity considered in this paper is described by the following action:

\be \label{action}
S = \int d^4 x \sqrt{-g } \left[ \frac{R}{ 16 \pi}  + \Lambda^4 \mathcal{F}( X, W^{ij}) \right]
\ee
Here $R$ is the scalar curvature and $\mathcal{F}$  is a function of scalar fields $\Phi^{0}, \Phi^{i}$. The functions $X$ and $W^{i j}$ are defined as,
\be
X = \frac{\partial^{\mu} \Phi^0 \partial_{\mu} \Phi^0 }{ \Lambda^4}
\ee
\be
W^{i j} = \frac{\partial^{\mu} \Phi^i \partial_{\mu} \Phi^j}{ \Lambda^4} - \frac{\partial^{\mu} \Phi^i \partial_{\mu} \Phi^0  \partial^{\nu} \Phi^j \partial_{\nu} \Phi^0 }{ \Lambda^4 X}
\ee
Here $\Lambda$ has dimensions of mass. The theory described by action in eq.$\refb{action}$ is the low-energy effective theory below the ultra-violet cutoff $\Lambda$. Perturbative analysis on the theory calculate the value of $\Lambda$ to be in the order of $ \sqrt{ m M_{pl}}$ where $m$  is the graviton mass and $M_{pl}$ the Plank mass\cite{luty}  \cite{dubo} \cite{ruba} \cite{pilo1}. The scalar fields $\Phi^0, \Phi^i$ are minimally coupled to gravity;  these fields are responsible for spontaneously breaking Lorentz symmetry. When  symmetry is broken, the scalar fields $\Phi^0, \Phi^i$ acquire a vacuum expectation value which depends on space-time. These fields are called Goldstone fields.

Given the fact that black holes are a part of our universe, studies of black holes and their stability properties in massive gravity theories are well founded. In this paper we study static spherically symmetric black hole solutions to the theory described above. In particular, we study the QNM frequencies and stability properties of a scalar field around these black holes. The paper is organized as follows: In section 2, an introduction to black holes in massive gravity is given. In section 3, the scalar field perturbations are introduced. In section 4, the QNM frequencies are computed. In section 5 the null geodesics approach is used to compute QNM frequencies. In section 6, the P$\ddot{o}$schl-Teller approximation is used to computed QNM frequencies for large $l$ and large $\lambda$. In section 7, the massive scalar field perturbation is studied. Finally, in section 8, the conclusion is given.


\section { Introduction to black holes in massive gravity }

In this section we will present static spherically symmetric  black hole solutions to the action in eq.$\refb{action}$. We will not derive it here since a detailed derivation is given in \cite{tinya} and \cite{pilo}.

The metric for the solution is given by,
\begin{equation} \label{metric}
ds^2 = - f(r) dt^2 + \frac{ dr^2}{ f(r)} + r^2 ( d \theta^2 + sin^2 \theta d \phi^2)
\end{equation}
where,

\begin{equation}
f(r) = 1 - \frac{ 2 M} { r} - \frac{ Q}{r^{\lambda}}
\end{equation}
The scalar fields are given by,
\be
\Phi^0 = \Lambda^2 \left( t + h(r) \right); \hspace{1 cm} \Phi^i =   \Lambda^2 x^i
\ee
where
\be
h(r) = \pm \int \frac{ dr} { \alpha(r)} \left[ 1 - \alpha(r) \left( \frac{ Q \lambda(\lambda-1)}{ 12 m^2}\frac{1}{ r^{\lambda+2}} + 1 \right)^{-1} \right]^{1/2}
\ee
Here $m$ is the mass of the graviton  and $\lambda$ is a positive constant. In order for the solutions to be asymptotically flat and $M$ to be the ADM mass, $\lambda$ has to be greater than 1.  $Q$ is a scalar charge and represents the modifications to Einstein's general relativity due to the presence of graviton with a mass.\\
\noi

The function $\mathcal{F}$ for this particular solution is given by,
\be
\mathcal{F} = \frac{ 12} { \lambda} \left( \frac{ 1}{ X} + w_1 \right) - \left( w_1^3 - 3 w_1 w_2 - 6 w_1 + 2 w_3 - 12\right)
\ee
where,
\be
w_n = Tr( W^n)
\ee

There are two possibilities for $Q$ when $M >0$.

\subsection{ $Q >0$}
The function $f(r)$ for this case is plotted in Fig.1.  The geometry is similar to the Schwarzschild black hole with a single horizon. The event horizon $r_h$ is always larger than the one for the Schwarzschild black hole. This is shown in the Fig.2.

\begin{center}
\scalebox{.9}{\includegraphics{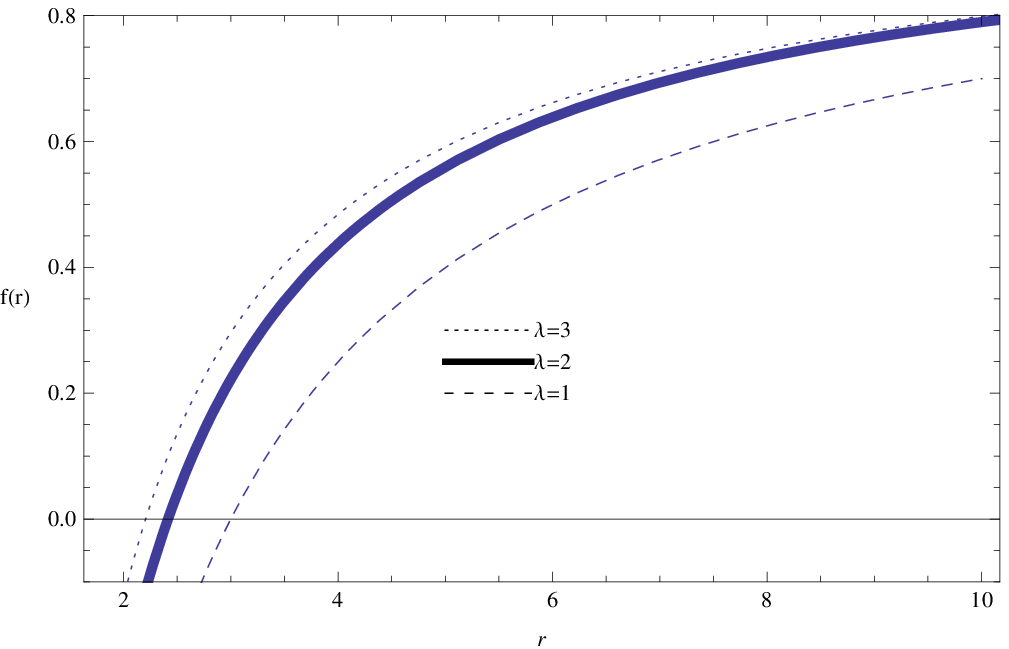}}
\vspace{0.3cm}
 \end{center}

Figure 1. The figure shows the $f(r)$ vs $r$. Here, $M =1$ and  $Q = 1$.\\

The horizon size decreases with the value of $\lambda$ as shown in Fig. 2.

\begin{center}
\scalebox{.9}{\includegraphics{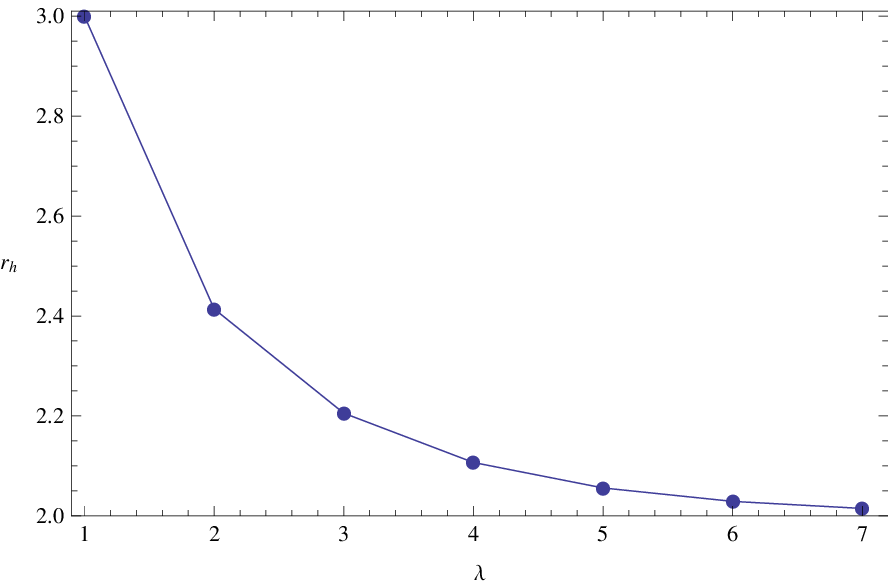}}
\vspace{0.3cm}
 \end{center}

Figure 2. The figure shows the $r_h$ vs $\lambda$. Here, $Q = 1$ and $M = 1$.\\

For a test particle  in the vicinity of the black hole, the effective potential for zero angular momentum is given by $ V = f(r)$. The force $F(r) = -\frac{1}{2} \frac{dV}{dr} $. The potential is plotted in Fig.3 and the corresponding force is plotted in Fig.4. One can observe that the force is attractive all times. However, the force is greater for the massive gravity black hole in comparison with the Schwarzschild black hole.

\begin{center}
\scalebox{.9}{\includegraphics{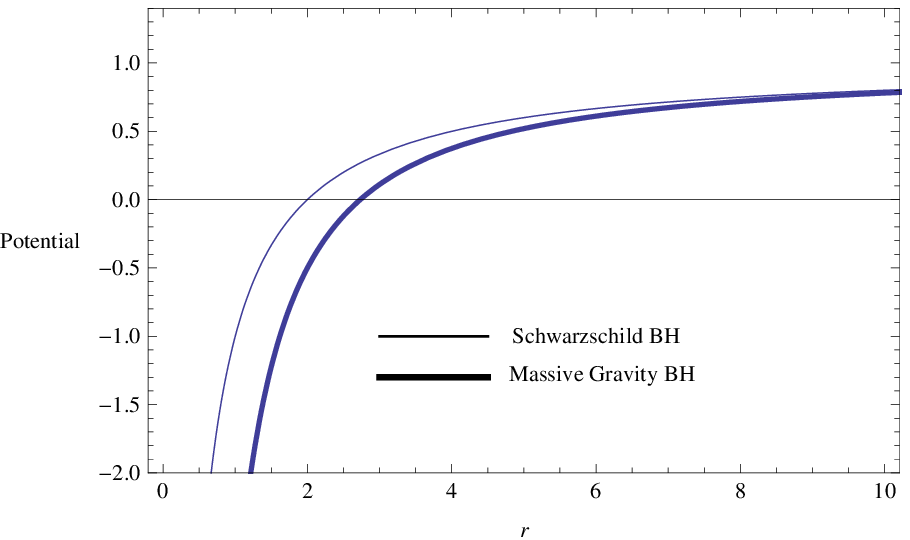}}
\vspace{0.3cm}
 \end{center}

Figure 3. The figure shows the $V$ vs $r$ for a test particle. Here, $Q = 1$ and $M = 1$.\\

\begin{center}
\scalebox{.9}{\includegraphics{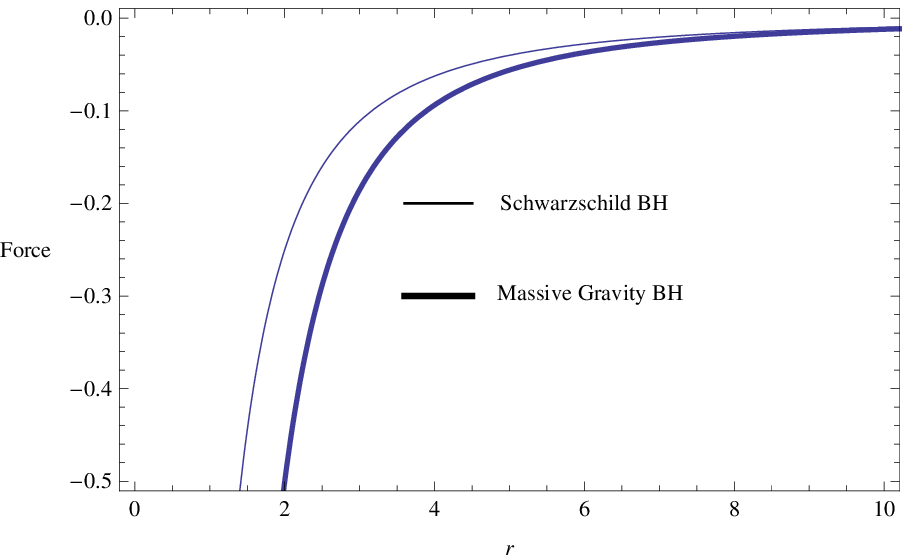}}
\vspace{0.3cm}
 \end{center}

Figure 4. The figure shows the $Force$ vs $r$ for a test particle. Here, $Q = 1$ and $M = 1$.\\


\subsection{ $ Q < 0$}

In this case, there are three possibilities.  There could be two, single or no horizons depending on the relation between the parameters. The critical mass which separate these three cases are given by,
\be
M_{critical} = \frac{ \lambda |Q|^{1/\lambda}}{ 2} \left( \frac{ 1}{ \lambda -1} \right)^{\frac{ \lambda -1}{ \lambda}} 
\ee
When $M > M_{critical}$, there will be two horizons. When $ M = M_{critical}$, there will be degenerate horizons. When $M < M_{critical}$, there won't be horizons and it will be a naked singularity. The geometry is very similar to the Reissner-Nordstrom black hole. The function $f(r)$ for all three cases are  represented in Fig.5.

\begin{center}
\scalebox{.9}{\includegraphics{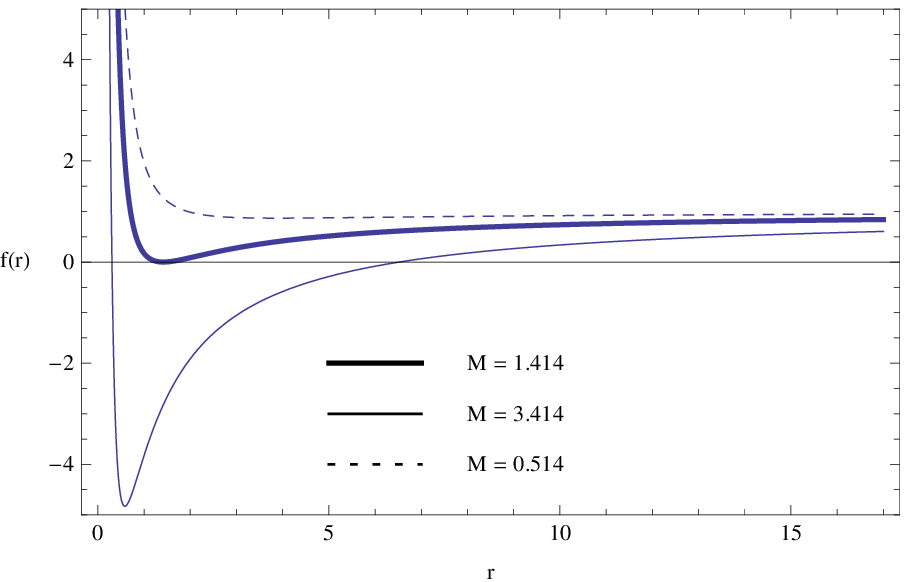}}
\vspace{0.3cm}
 \end{center}

Figure 5. The figure shows  $f(r)$ vs $r$. Here, $Q = -2$ and $\lambda = 2$.\\

For a test particle  in the vicinity of the black hole, the effective potential for zero angular momentum is given by $ V = f(r)$. The force $F(r) =  -\frac{1}{2}\frac{dV}{ dr} $. The potential is plotted in Fig.6 and the corresponding force is plotted in Fig.7.

\begin{center}
\scalebox{.9}{\includegraphics{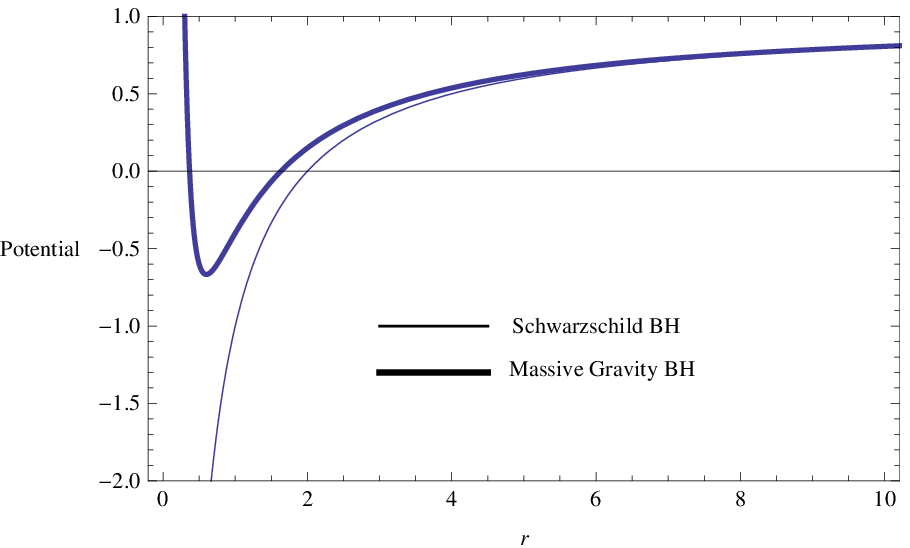}}
\vspace{0.3cm}
 \end{center}

Figure 6. The figure shows the $V$ vs $r$ for a test particle. Here, $Q = -0.6$ and $M = 1$.\\

\begin{center}
\scalebox{.9}{\includegraphics{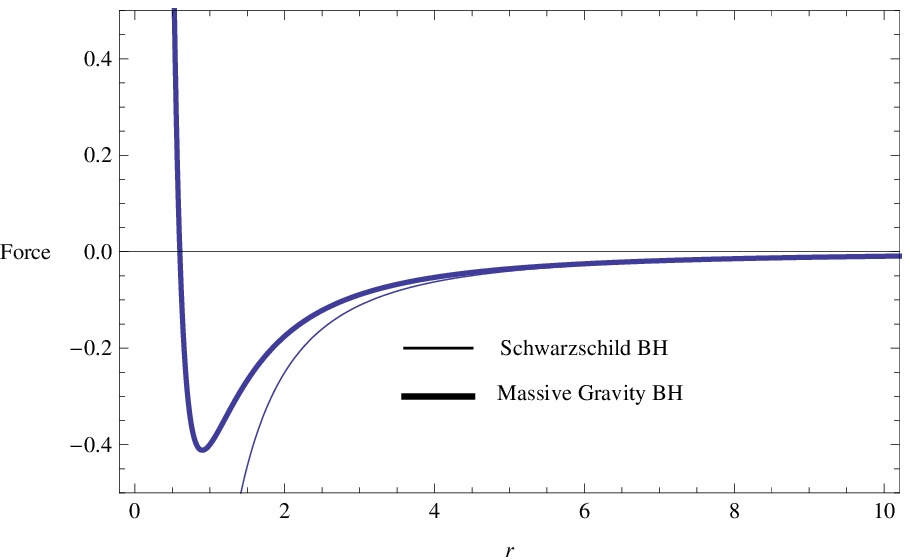}}
\vspace{0.3cm}
 \end{center}

Figure 7. The figure shows the $Force$ vs $r$ for a test particle. Here, $Q = -0.6$ and $M = 1$.\\

As shown in the Fig.7, the force is attractive for the massive gravity only up to a certain distance. After that the force is repulsive.\\

In the rest of the paper, we will focus on  $Q >0$ case only.

\subsection{ Temperature}

 The Hawking temperature  of the black hole is given by,
 
 \be
 T_H =  \frac{ 1}{ 4 \pi}  \left| \frac{ df(r)}{ dr} \right|_ { r = r_h} = \frac{ 2 M}{ r_h} + \frac{ Q \lambda}{ r_h^{ \lambda + 1}}
 \ee

Temperature vs M is plotted in Fig.8. When  M increases, the temperature decreases which is similar to the Schwarzschild black hole.  Temperature vs Q is plotted in Fig.9. When Q is increased, the temperature increases and then decreases. Thermodynamics and phase structure of the massive gravity black hole studied in this paper was addressed by Capela and Nardini 
\cite{nar}.

 \begin{center}
\scalebox{.9}{\includegraphics{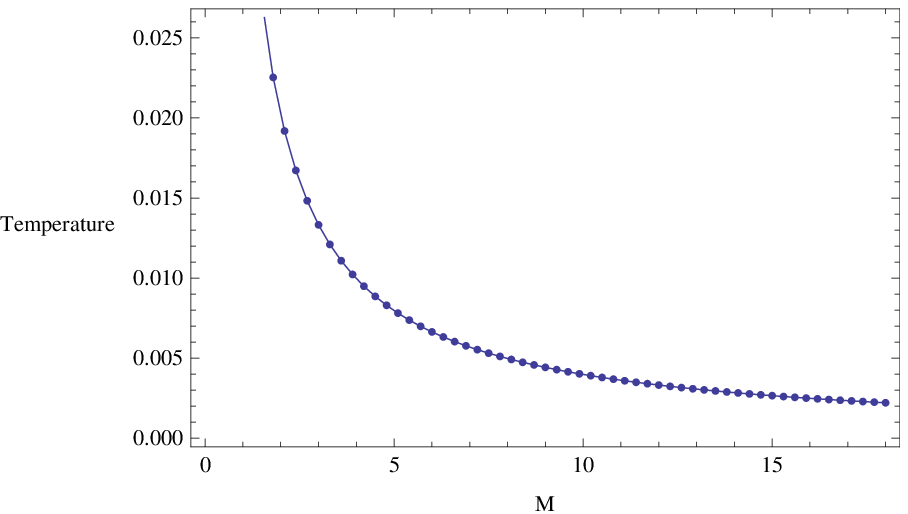}}
\vspace{0.3cm}
 \end{center}

Figure 8. The figure shows the $Temp$ vs $M$.  Here $ Q=1$ and $\lambda =3$.\\

 \begin{center}
\scalebox{.9}{\includegraphics{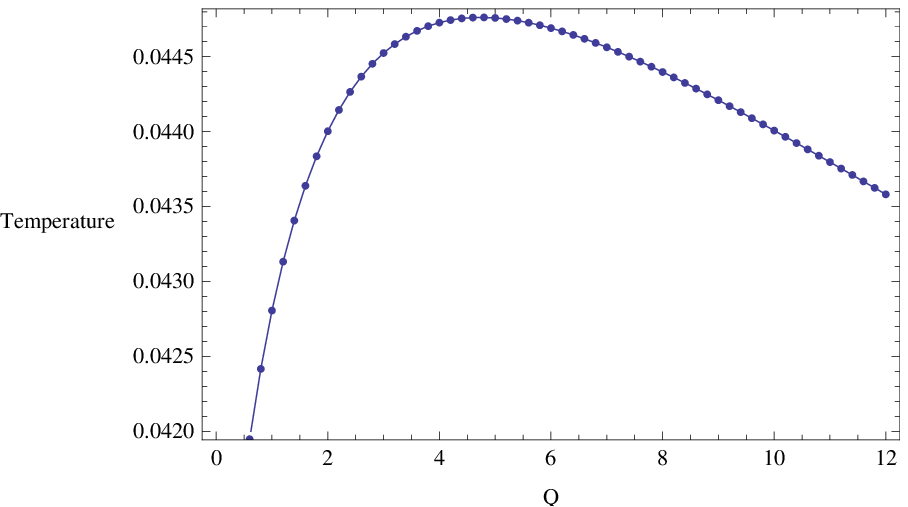}}
\vspace{0.3cm}
 \end{center}

Figure 9. The figure shows the $Temp$ vs $Q$.  Here $ M=1$ and $\lambda =3$.\\


\section{ Basic equations for the scalar perturbations of the black hole in massive gravity}

Klein-Gordon equation for the massless scalar field around the black hole in massive gravity is given by,
\be \label{klein}
\bigtriangledown^2 \psi = 0 
\ee
By separation of variables with the substitution,
\be
\psi = e^{ - i \omega t}  Y_{ l, m} ( \theta, \phi) \frac{ R(r)}{ r}
\ee
Simplifies the eq.$\refb{klein}$ leading to a Schr$\ddot{o}$dinger-type equation given by,
\be \label{wave}
\frac{ d^2 R(r_*) }{ dr_*^2} + \left( \omega^2 - V_{eff}(r_*)  \right) R(r_*) = 0
\ee
Here, $V_{eff}(r_*)$ is given by,
\be
V_{eff}(r_*) = \frac{ l ( l + 1) f(r)} { r^2} + \frac{ 1}{ 2 r}  \frac{ d( f(r)^2)}{dr}
\ee
$\omega$ is the frequency of the wave mode and $ Y_{ l,m}( \theta, \phi) $ is the spherical harmonics. $r_*$ is the tortoise coordinates which is given by,
\be
dr_* = \frac{ dr} { f(r)}
\ee
The effective potential $V_{eff}(r)$ depends on four parameters: $l, M, Q$ and $\lambda$. In Fig.10, $V_{eff}(r)$ is plotted as a function of $r$ by varying $l$. When $l$ increases, the height of the potential increases.
\begin{center}
\scalebox{.9}{\includegraphics{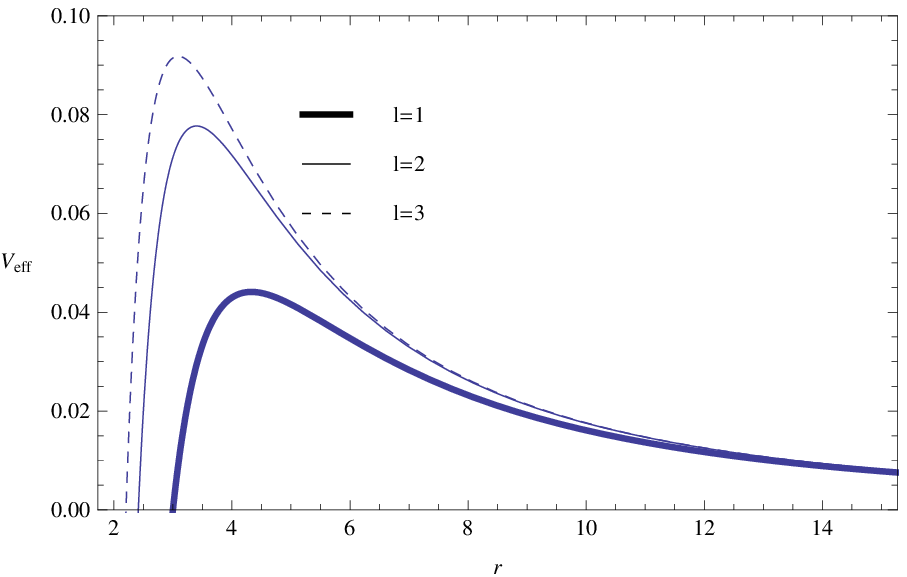}}
\vspace{0.3cm}
 \end{center}

Figure 10. The figure shows the $V_{eff}(r)$ vs $r$.  Here $ M =1, Q =1$ and $\lambda = 1$.\\

In Fig.11  $V_{eff}(r)$ is plotted as a function of $r$ by varying $M$. When mass increases, the height of the potential decreases.

\begin{center}
\scalebox{.9}{\includegraphics{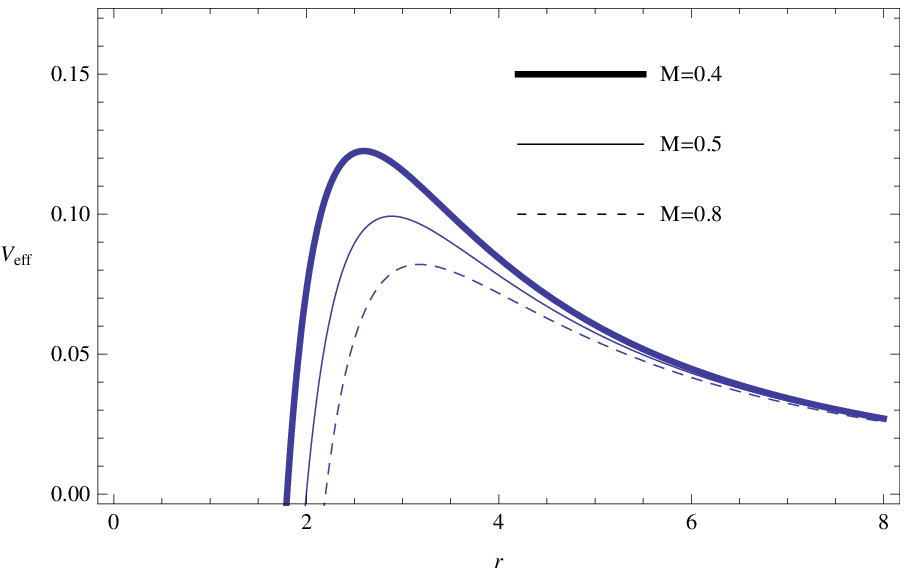}}
\vspace{0.3cm}
 \end{center}

Figure 11. The figure shows the $V_{eff}(r)$ vs $r$.  Here $ Q =1, \lambda = 1$ and $ l = 1$.\\

In Fig.12 , $V_{eff}(r)$ is plotted as a function of $r$ by varying $Q$. When $Q$ increases, the potential height decreases.

\begin{center}
\scalebox{.9}{\includegraphics{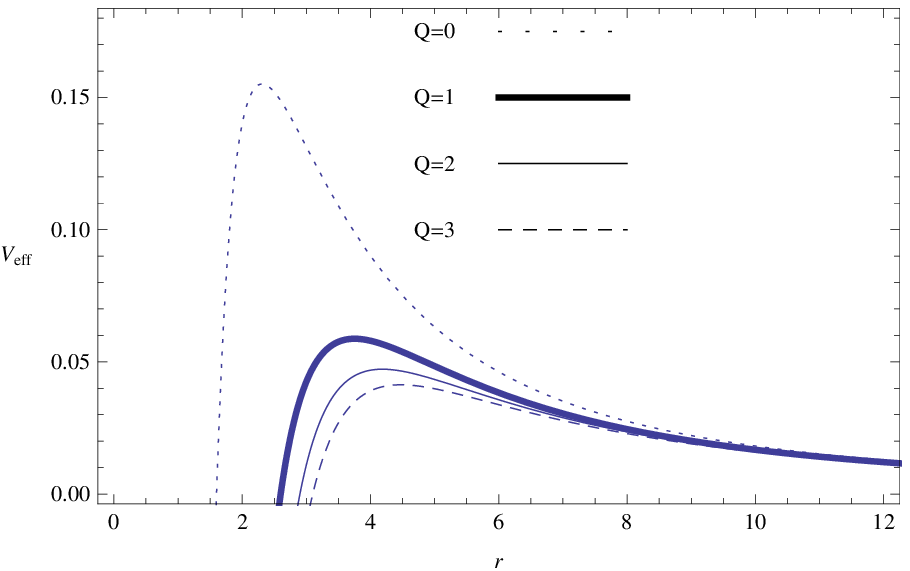}}
\vspace{0.3cm}
 \end{center}

Figure 12. The figure shows the $V_{eff}(r)$ vs $r$.  Here $ M =0.8, l =1$ and $\lambda = 1$.\\

 In Fig. 13, $V_{eff}(r)$ is plotted as a function of $r$ by varying $\lambda$. When $\lambda$ increases, the height of the potential increases.

\begin{center}
\scalebox{.9}{\includegraphics{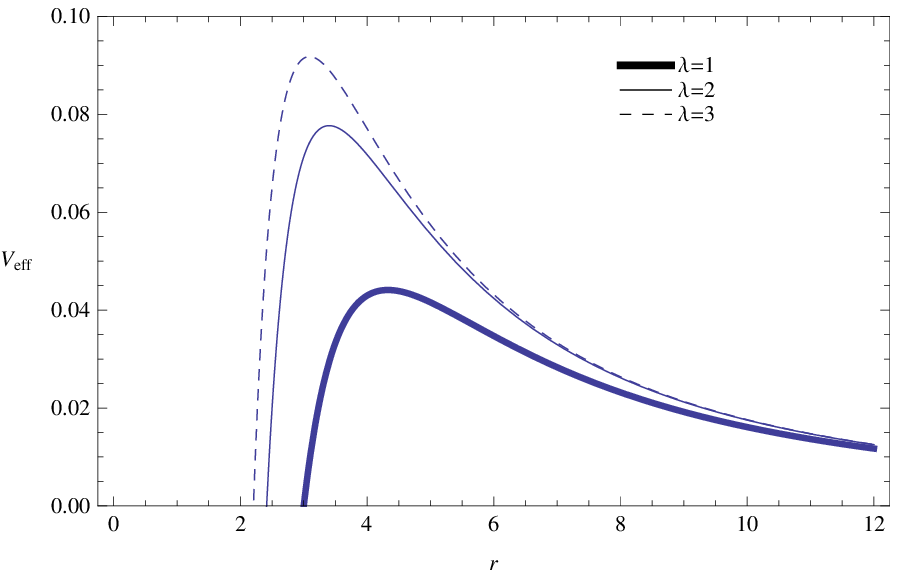}}
\vspace{0.3cm}
 \end{center}

Figure 13. The figure shows the $V_{eff}(r)$ vs $r$.  Here $ M =1, Q =1$ and $l = 1$.\\


\section{ Quasi-normal modes of massive gravity black hole}

Quasi-normal modes (QNM) are the solutions to the wave equation in eq.$\refb{klein}$. For an asymptotically flat black hole, such as the one considered here, the boundary conditions are, ingoing waves at the horizon and out going waves at the spatial infinity. QNM frequencies are complex numbers with $\omega$ having a real part ($\omega_R$) and an imaginary part  ($\omega_I$). Since the time dependence of the wave solution goes as $ e^{- i \omega t}$, for stable solutions, one expects $\omega_I$ to be negative.

QNM frequencies are labeled by an integer $n$. $ n =0$ corresponds to the fundamental mode and it will have the smallest value of $\omega_I$. The overtones with high $n$ will have larger values of $\omega_I$. Hence $\omega_I( n=0)$ will decay slower than the higher over tones. Hence in this paper we will be particularly interested in the fundamental mode frequency.

We will use the WKB approach developed by Iyer and Will \cite{will}  and later extended to sixth order by Konoplya \cite{kono4}.  In this method, the QNM frequencies are given by,
\be
i \frac{ \omega^2 - V_m }{ \sqrt{ - 2 V_{m}''}} = \Lambda _2 + \Lambda_3 + \Lambda_4 + \Lambda_5 + \Lambda_6 + n + \frac{1}{2}
\ee
Here, $V_m$ is the maximum value of the potential. $V_m''$ is the value of the second derivative of the effective potential at the place where it is maximum.  The specific expressions for $\Lambda_i$ can be found in \cite{kono4} and \cite{will}. 6th order WKB approach has been applied to find $\omega$ for   the Bardeen black hole \cite{fernando1}  and for the Born-Infeld black hole \cite{fer2}.

In Fig.14, $\omega$ is plotted against $\lambda$.  For small values of $\lambda$, both $\omega_R$  and $\omega_I$  increases. When $\lambda$ gets larger, $\omega_R$ converges to a stable value.  When $\lambda$ is increased, $\omega_I$ reaches a maximum before decreasing to a stable value. The stable value reached is the $\omega$ for the Schwarzschild black hole with the same mass and $l$ which is $0.241821- i 0.048383$.
Hence the massive gravity black hole decays slower than the Schwarzschild black hole except for a range of  values of $\lambda$ which is evident from the graph. This would be expected since for large $\lambda$, the function $f(r)$ behaves very similar to the function $f(r)$ for the Schwarzschild black hole which is $( 1- \frac{2 M}{r})$.

\begin{center}
\scalebox{.9}{\includegraphics{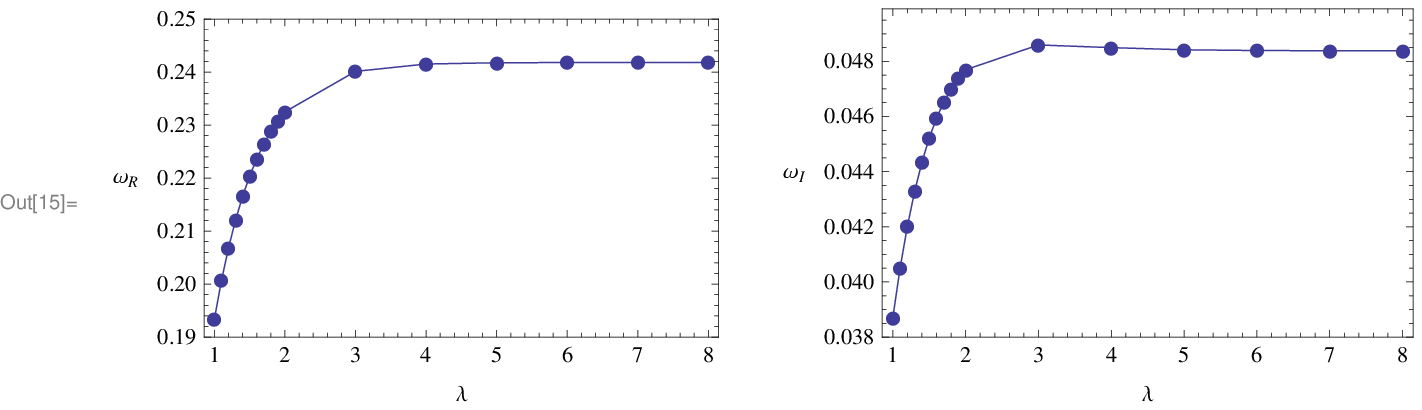}}
\vspace{0.3cm}
 \end{center}

Figure 14. The figure shows the $\omega$ vs $\lambda$. Here $ M = 2, Q =1$ and $ l = 2$.\\

In Fig.15, $\omega$ is plotted against the scalar charge $Q$. $\omega_R$  decreases and $\omega_I$  increases with $Q$. Hence for large $Q$, the modes decays faster. Compared to the Schwarzschild black hole, which corresponds to $Q=0$ case, the modes decays faster.

\begin{center}
\scalebox{.9}{\includegraphics{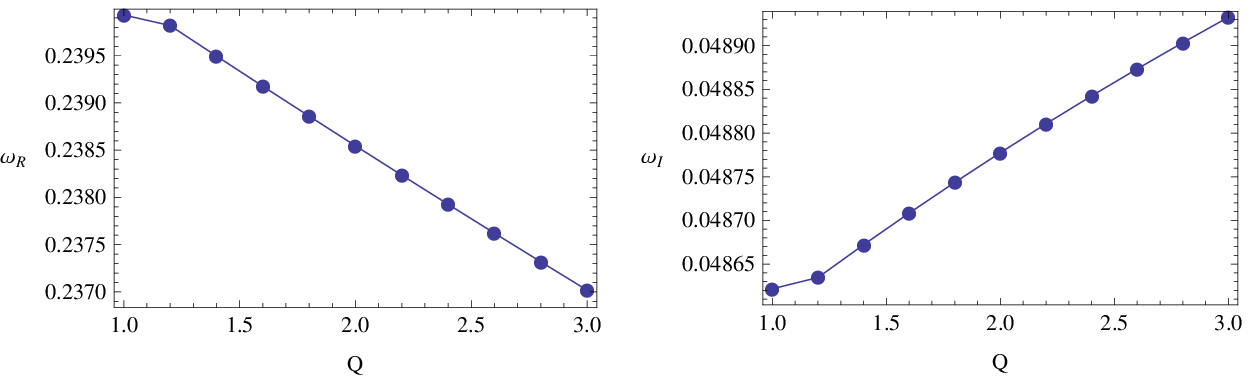}}
\vspace{0.3cm}
 \end{center}

Figure 15. The figure shows the $\omega$ vs $Q$.  Here $ M =2, \lambda = 3$ and $ l =2$.\\

In Fig.16, $\omega_R$ is plotted against the mass of the black hole, $M$. $\omega_R$  decreases with $M$.  Compared to the Schwarzschild black hole, which corresponds to $Q=0$ case, $\omega_R$ is smaller for the massive gravity black hole.

\begin{center}
\scalebox{.9}{\includegraphics{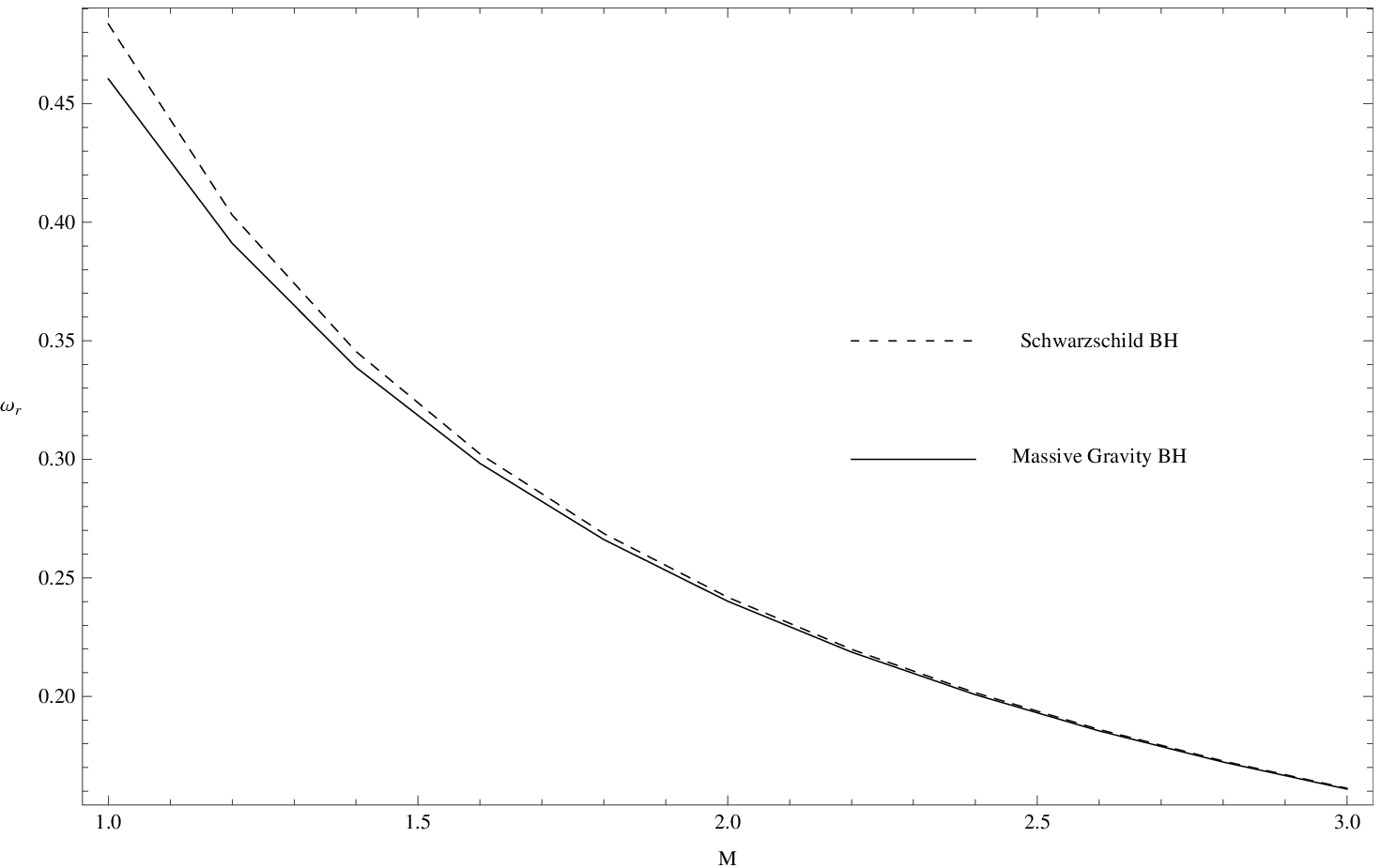}}
\vspace{0.3cm}
 \end{center}

Figure 16. The figure shows the $\omega_R$ vs $M$. Here $ Q =1, \lambda =3$ and $ l =2$.\\

In Fig.17, $\omega_I$ is plotted against the mass of the black hole, $M$. $\omega_I$  decreases with $M$. Hence for large $M$, the modes decays slower. Compared to the Schwarzschild black hole, which corresponds to $Q=0$ case, $\omega_I$ is larger for the massive gravity black hole. Hence, the scalar field decays faster in the massive gravity black hole with the same mass. 

\begin{center}
\scalebox{.9}{\includegraphics{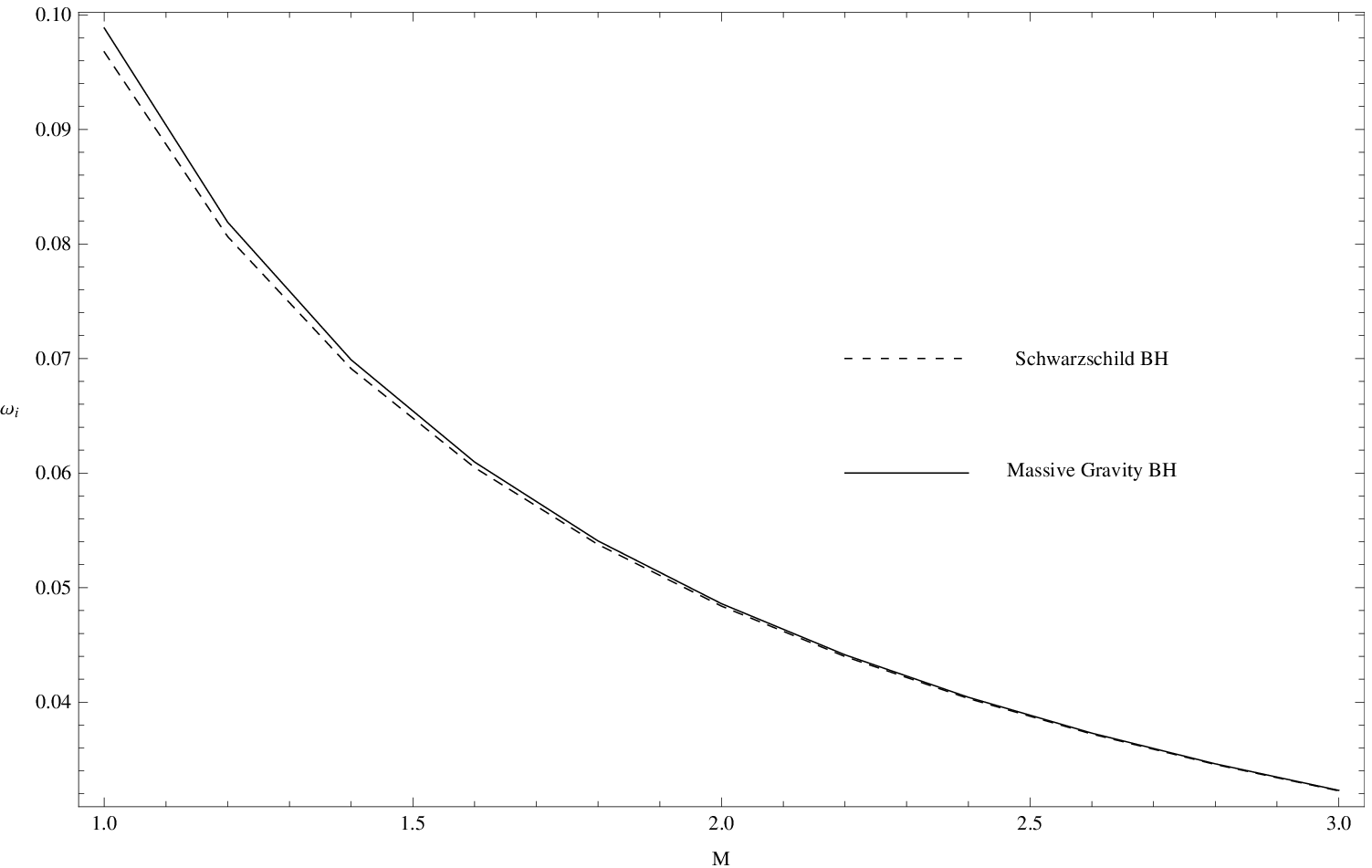}}
\vspace{0.3cm}
 \end{center}

Figure 17. The figure shows the $\omega_I$ vs $M$. Here $ Q =1, \lambda =3$ and $ l =2$.\\

In Fig.18, $\omega_R$ is plotted against $l$ for both $n=0$ and $ n=1$. Both cases shows a linear dependency on $l$.

\begin{center}
\scalebox{.9}{\includegraphics{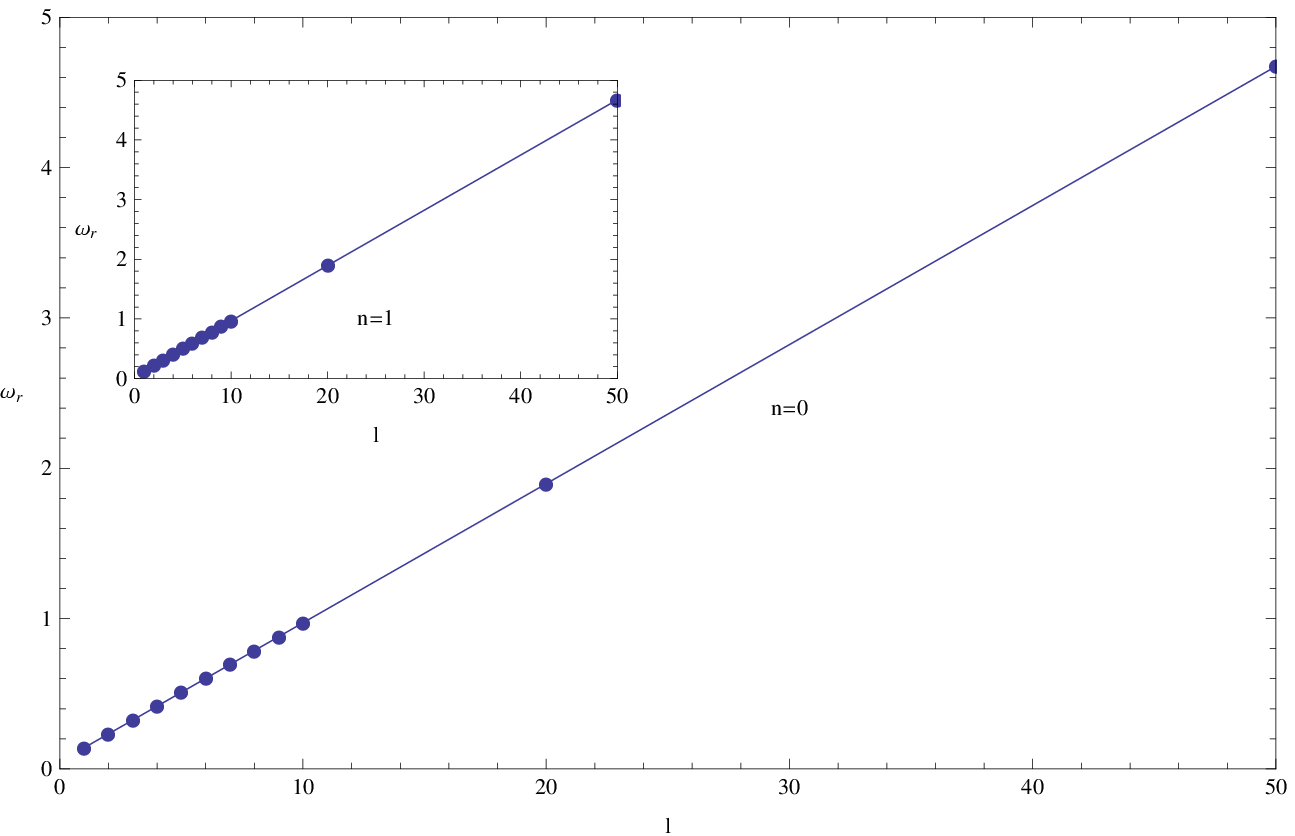}}
\vspace{0.3cm}
 \end{center}

Figure 18. The figure shows the $\omega_R$ vs $l$. Here $ M = 2, Q =1$ and $ \lambda =2$.\\

In Fig.19, $\omega_I$ is plotted against $l$ for both $n=0$ and $ n=1$.  $\omega_I$ decreases as $l$ increases and then approach a constant value.

\begin{center}
\scalebox{.9}{\includegraphics{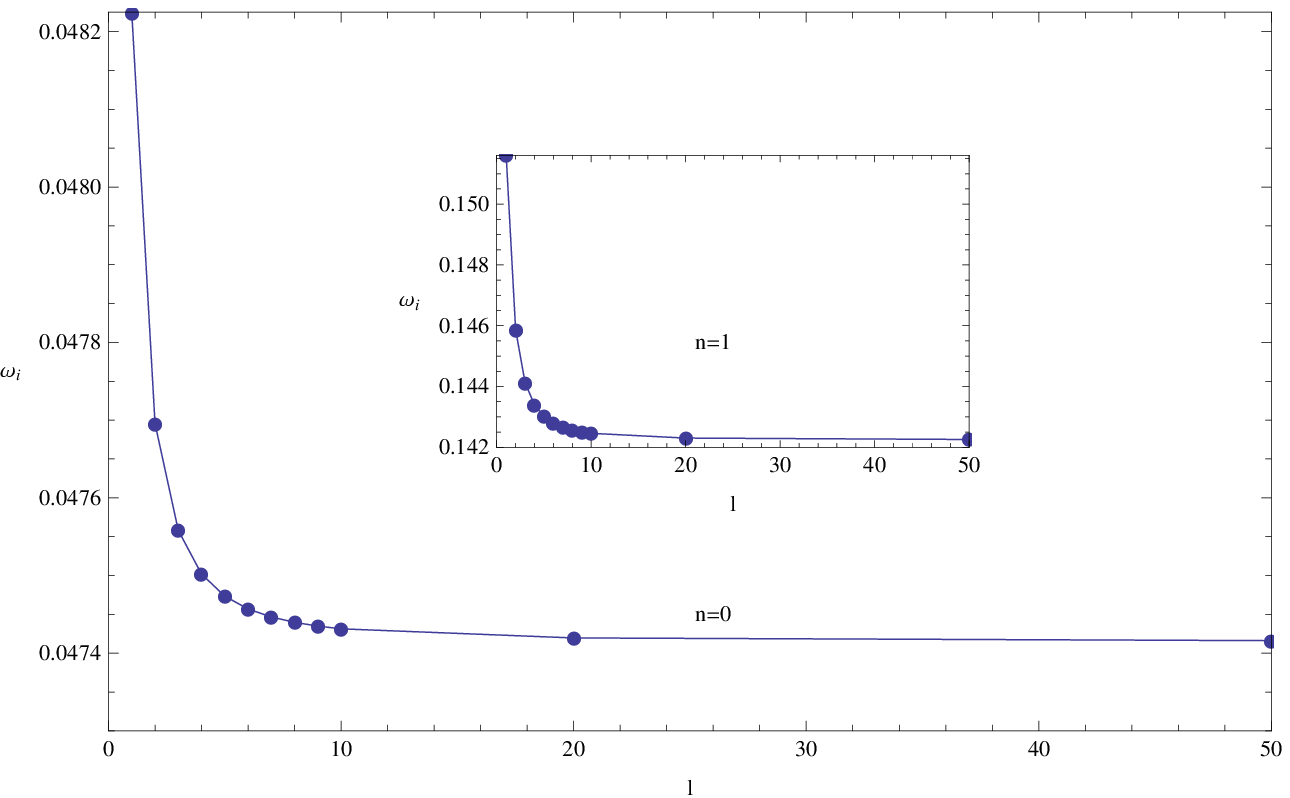}}
\vspace{0.3cm}
 \end{center}

Figure 19. The figure shows the $\omega_I$ vs $l$. Here $ M = 2, Q =1$ and $ \lambda =2$.\\


\section{QNM's and the  null geodesics of the black hole}

Studies of null geodesics takes a prominent place in understanding  the properties of a black hole. Null geodesics, which gives the path of a massless particle around a black hole can facilitate  computing  QNM frequencies. If the black hole has an unstable null geodesic, which is the case for the one considered here, the QNM frequencies at the eikonal limit ( $l >>  1$) can be computed via the properties  of the null geodesics. This method was presented by Cardoso et.al. in \cite{cardoso}. To give an explanation of this approach, let us first present the equation for the null geodesics as,

\be
\dot{r}^2 = E^2 - V_{null}
\ee
Here, $V_{null}$ is the effective potential for the null  geodesic given by,
\be
V_{null} = \frac{L^2}{r^2} f(r)
\ee
Here, $L$ is the angular momentum of the massless particle describing the null geodesics.  Fig. 20 presents the plot of $V_{null}$ vs $r$. One can observe that the potential for the massless scalar field and the potential for the null geodesics are similar in shape. In fact,  when $ l \ra \infty$, the maximum of the scalar field potential and the maximum of the null geodesics occur at the same value of $r$. The QNM frequencies at $ l \ra \infty$ is given by \cite{cardoso},

\begin{center}
\scalebox{.9}{\includegraphics{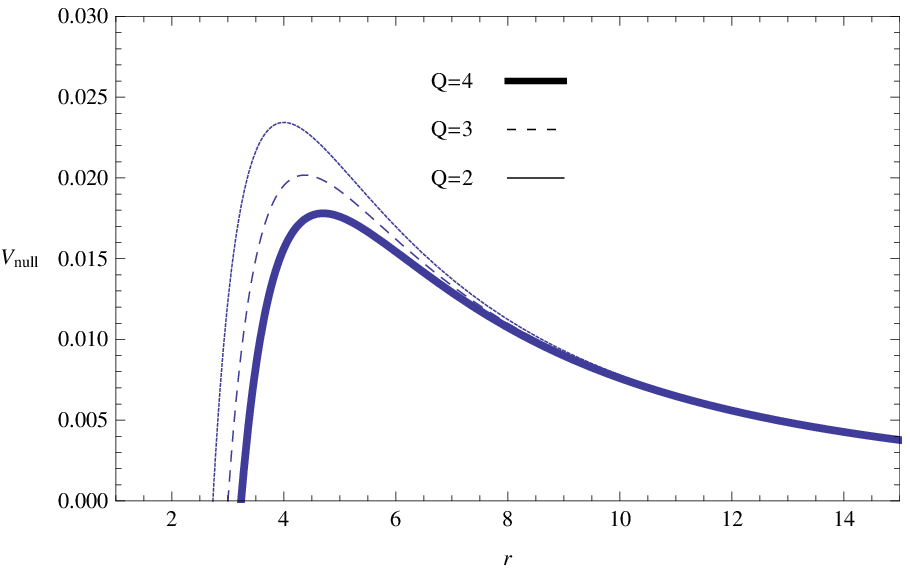}}
\vspace{0.3cm}
 \end{center}

Figure 20. The figure shows the $V_{null}$ vs $r$.  Here $ M =1$ and $\lambda =2$

\begin{equation}
\omega_{QNM} = \Omega_c l - i ( n + \frac{1}{2} ) | \Lambda_c|
\end{equation}
Here, $\Omega_c$ is the coordinate angular velocity of the massless particle given as,
\begin{equation}
\Omega_c = \frac{ \dot{\phi(r_c)}}{\dot{t}(r_c)}  = \sqrt{ \frac{ f(r_c)}{r_c^2 }} 
\end{equation}
and $\Lambda_c$ is the Lyapunov exponent  which is the decay rate of the massless particle at the circular geodesics given by,
\begin{equation}
\Lambda_c = \sqrt{ \frac{-V_{null}''(r_c) r_c^2 f(r_c)}{2 L^2} }
\end{equation}
In Fig. 22, $\Omega_c$ is plotted which represents the real part of the QNM frequency. If one compare this with the Fig. 18, where WKB approach values are plotted, it shows similar behavior. On the other hand, in Fig. 21,  $\Lambda_c$ is plotted. Since $\Lambda_c$ is proportional to $\omega_I$, one can compare the behavior with the plot in Fig. 18.   $\omega_I$ increases for the initial values of $Q$ similar to the one in Fig.18. However, it seems $\omega_I$ decreases for large values of $Q$ as evident from the Fig. 21.

\begin{center}
\scalebox{.9}{\includegraphics{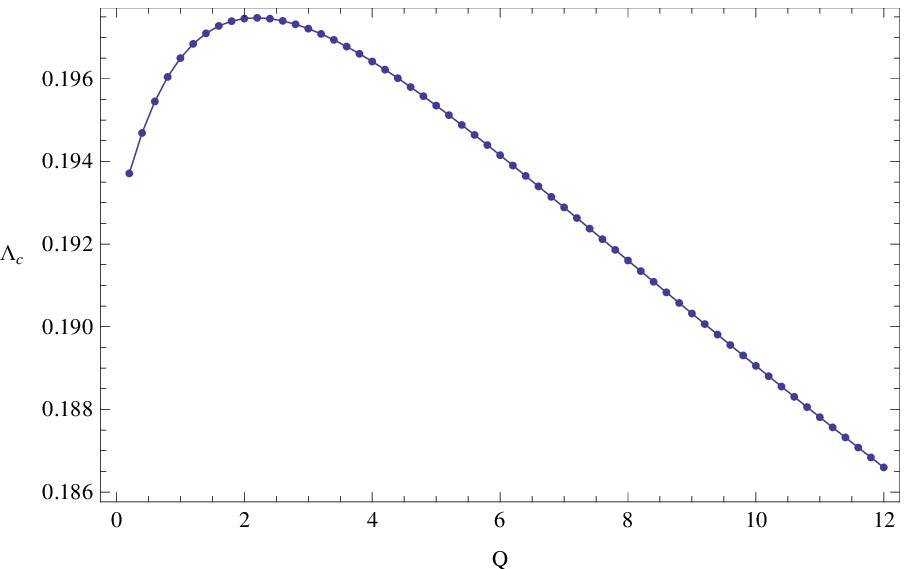}}
\vspace{0.3cm}
 \end{center}

Figure 21. The figure shows the $\Lambda_c$ vs $Q$.  Here $ M =1$ and $\lambda =3$\\

\begin{center}
\scalebox{.9}{\includegraphics{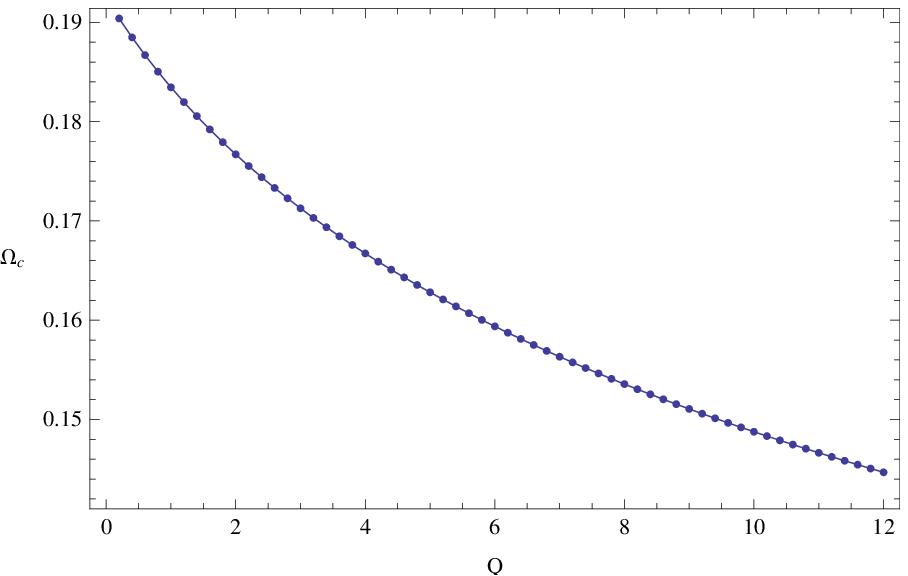}}
\vspace{0.3cm}
 \end{center}

Figure 22. The figure shows the $\Omega_c$ vs $Q$. Here $ M=1$ and $ \lambda =3$\\

An interesting paper relating black hole spectroscopy with null geodesics is given by Wei et.al. in \cite{wei2}.


\section{ P$\ddot{o}$schl-Teller approximation leading to analytical $\omega$}

Exact solutions to the wave equation eq.$\refb{wave}$ are rare. There are very few cases where exact solutions are available in the literature. The BTZ black hole \cite{cardoso2} and the charged dialton black hole in 2+1 dimensions \cite{fernando6} are two of such cases that comes to the minds of the authors. 

Ferrari and Mashhoon \cite{mash}, introduced a new analytical approach to find QNM frequencies by approximating the effective potential with the P$\ddot{o}$schl-Teller potential. They applied this method to obtain exact results for the Schwarzschild black hole, Reissner-Nordstrom black hole and the Kerr black hole. Since the P$\ddot{o}$schl-Teller potential can be solved exactly, one can find approximate analytical formulas for the QNM frequencies. Here, we will apply this method to obtain approximate expansions for QNM frequencies for the massive gravity black hole with large $\lambda$.

In  the  the P$\ddot{o}$schl-Teller approximation, the effective potential is approximated by the the potential given by,
\be
V = \frac{ V_o}{ Cosh^2 \alpha(r_* - r_{*o}) }
\ee
Here $r_*$ is the tortoise coordinate described in eq.(16) and $r_{*o}$ is the point where the potential reach the maximum. Hence, $\frac{dV}{dr_*} = 0$ at $r_* = r_{*o}$. The variable $\alpha$ is given by,
\be
\alpha^2 = \frac{1}{ 2 V_o} \frac{d^2V}{ dr_*^2} |_{r = r_{*o}}
\ee
where,
\be
V_o = V(r_* = r_{*o})
\ee
It was shown in \cite{cardoso2} \cite{mash} that the QNM frequencies for the above potential is given by,
\be
\omega = \pm \sqrt{ V_o - \alpha^2/4} - i \alpha ( n + 1/2)
\ee
Due to the complex nature of the potential in eq.(15), we will study the QNM frequencies at the eikonal limit (large $l$). Then, the dominant term in the effective potential is the one proportional to $l  ( l + 1)$. Hence,
\be
V(r) \approx \frac{ l(l+1) f}{r^2}
\ee
Since we are studying the wave equation for large $\lambda$, we will also approximate the function $f(r)$ as,
\be \label{poshf}
f(r) \approx  1 - \frac{ 2 M}{r} - \bar{Q} t
\ee
Here $t = \frac{1}{\lambda}$ and $\bar{Q}$ is chosen such that  $f$ will be dimensionally correct. Notice that when $\lambda \ra \infty$, $f \approx 1 - \frac{ 2 M}{r}$. Hence for large $\lambda$, the horizon gets closer to the horizon of the Schwarzschild black hole. This was observed in  Fig.2. Hence this approximation is reasonable for this computation and will facilitate the computation which otherwise would be quite complex.

Now, $\frac{dV}{dr_*} = \left(\frac{dV}{ dr}\right) \left( \frac{ dr} {dr_*}\right) = 0$ lead to the solutions,
\be
r_o = \frac{ 3 M}{ (1 - \bar{Q}t)} \Rightarrow  r_{*o} = \frac{ M( 3 - Log 4)}{ ( 1 - \bar{Q} t)^2}
\ee
The final results for $\alpha$ and $V_o$ are given as below:
\be
\alpha^2 = \frac{ ( 1 - \bar{Q}t)^4} { 27 M^2}
\ee
\be
V_o = l ( l+1) \frac{ ( 1 - \bar{Q} t)^3 }{ 27 M^2}
\ee
Hence $\omega_I$ and $\omega_R$ are given as,
\be
\omega_I =  \frac{ ( 1 - \bar{Q}t)^2}{ 3 \sqrt{3} M}
\ee
and
\be
\omega_R = \frac{ ( 1 - \bar{Q}t)^2}{ 3 \sqrt{3} M} \sqrt{ \frac{l( l+1)}{( 1- \bar{Q} t)} - \frac{1}{4}}
\ee
When $\lambda \ra \infty$, $\omega_R$ and $\omega_I$ reaches the value for the Schwarzschild black hole obtained by Ferrari and Mashhoon in \cite{mash}.

By replacing $t$ with $1/\lambda$, we have plotted the frequencies in the following figures.

\begin{center}
\scalebox{.9}{\includegraphics{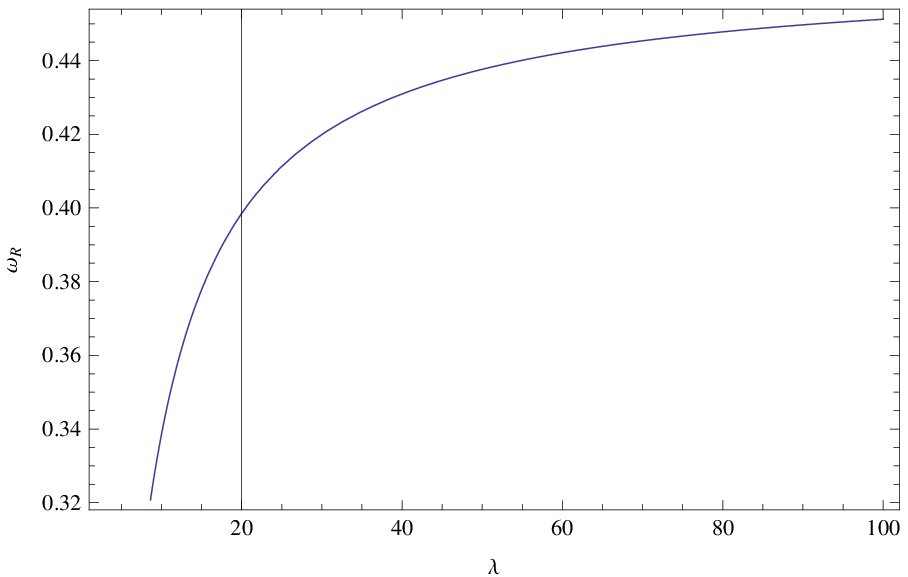}}
\vspace{0.3cm}
 \end{center}

Figure 23. The figure shows the $\omega_R$ vs $\lambda$. Here $ M=2$, $ l = 50$ and $Q =1$\\

\begin{center}
\scalebox{.9}{\includegraphics{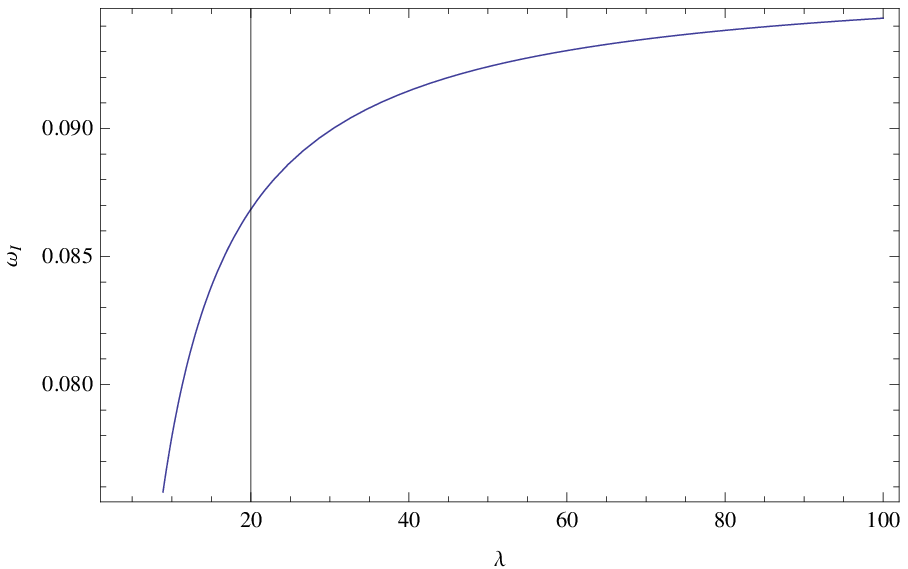}}
\vspace{0.3cm}
 \end{center}

Figure 24. The figure shows the $\omega_I$ vs $\lambda$. Here $ M=2$, $ l = 50$  and $ Q= 1$\\

Clearly, when $\lambda$ increases, $\omega_R$ and $\omega_I$  increases as shown in Fig.23   and Fig. 24 .  This behavior is similar to the behavior seen in Fig.14. Also, for large $\lambda$, both values of $\omega$ converges to the Schwarzschild black hole values similar to the behavior shown in Fig.14.  Hence the analytical approimation further verify the behavior of $\omega$ values obtained via a numerical approach.


\section{ Massive scalar perturbations}

In this section, we will study the QNM frequencies of a massive scalar field. For the Schwarzschild black hole, it has been observed that the massive scalar field decay slower than the massless field \cite{kono5}. It has also been observed that $\omega_I$ will disappear when the mass of the field is increased to sufficiently large values leading to the phenomenon called ${\it quasiresonance}$. Such quasiresonance behavior has been observed for the Kerr black hole \cite{kono6} and for the Reissner-Nordstrom black hole \cite{ohashi}.\\

The equation of motion for the massive scalar field is given by,

\be \label{kleinmassive}
\bigtriangledown^2 \psi - m^2 \psi = 0 
\ee
After separation of variables, the radial component has the following equation:
\be
\frac{ d^2 R(r_*) }{ dr_*^2} + \left( \omega^2 - V^m_{eff}(r_*)  \right) R(r_*) = 0
\ee
Here, $V_{eff}(r_*)$ is given by,
\be
V^m_{eff}(r_*) = \frac{ l ( l + 1) f(r)} { r^2} + \frac{ 1}{ 2 r}  \frac{ d( f(r)^2)}{dr} + m^2 f(r)
\ee
The effective potential $V_{eff}^m$ is plotted  in Fig.25 by varying the mass $m$. When the mass is increased, the height of the potential increases as shown in the Figure.27.  Also at a critical value of the mass $m$, the potential ceases to have a maximum.

\begin{center}
\scalebox{.9}{\includegraphics{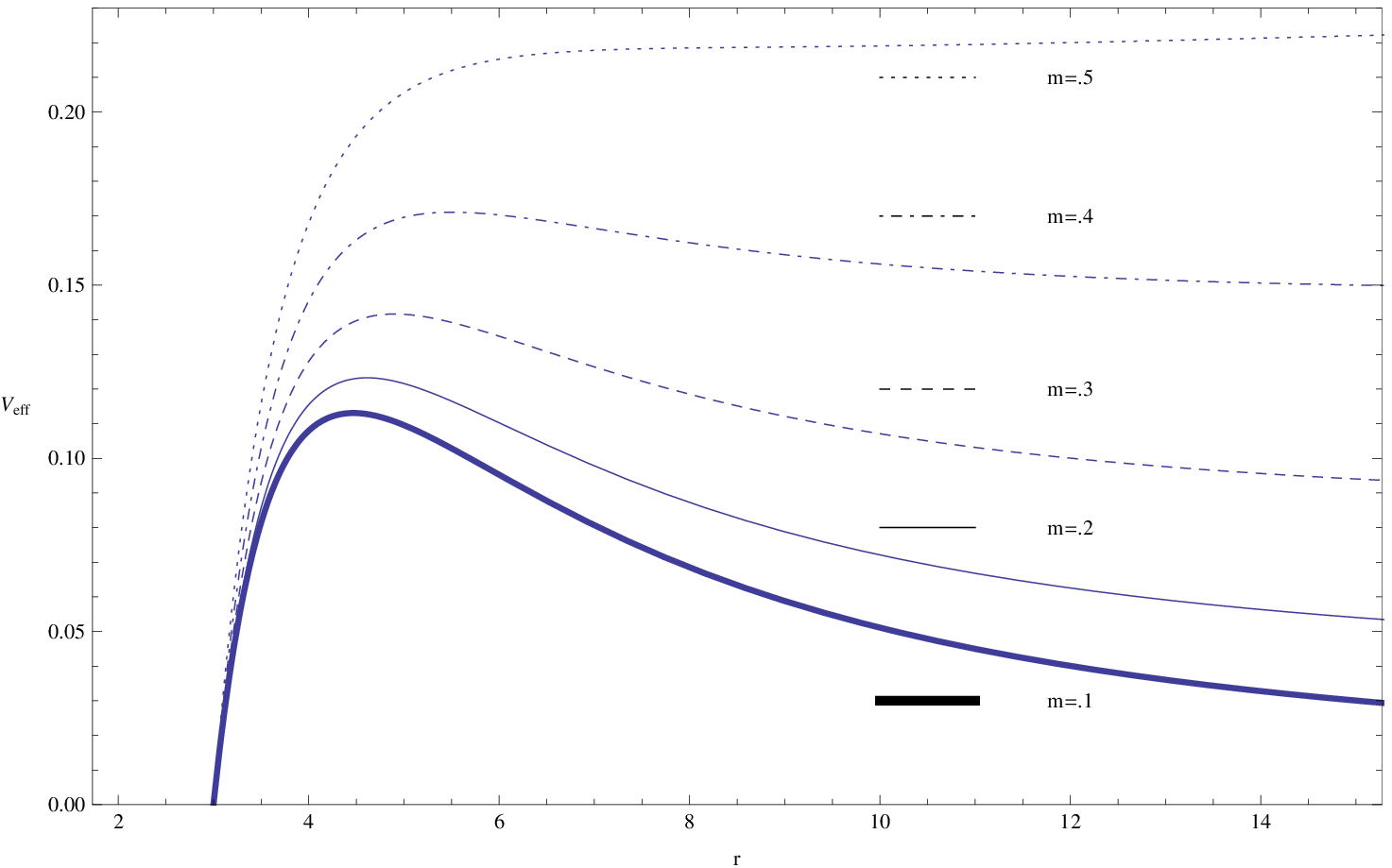}}
\vspace{0.3cm}
 \end{center}

Figure 25. The figure shows the $V^m_{eff}$ vs $r$.  Here $ M =1, Q =1$ and $l = 2$.\\

The QNM frequencies for the massive scalar field are computed using the 6th order WKB approach. Only the fundamental mode was computed. Fig. 26 shows  the real part of  $\omega$ for various values of mass $m$. We have computed  and plotted $\omega_R$ for the Schwarzschild black hole in the same plot for comparison. $\omega_R$ value increases with $m$ for both black holes. The Schwarzschild black hole has a higher value than the massive gravity black hole for a given mass. 

\begin{center}
\scalebox{.9}{\includegraphics{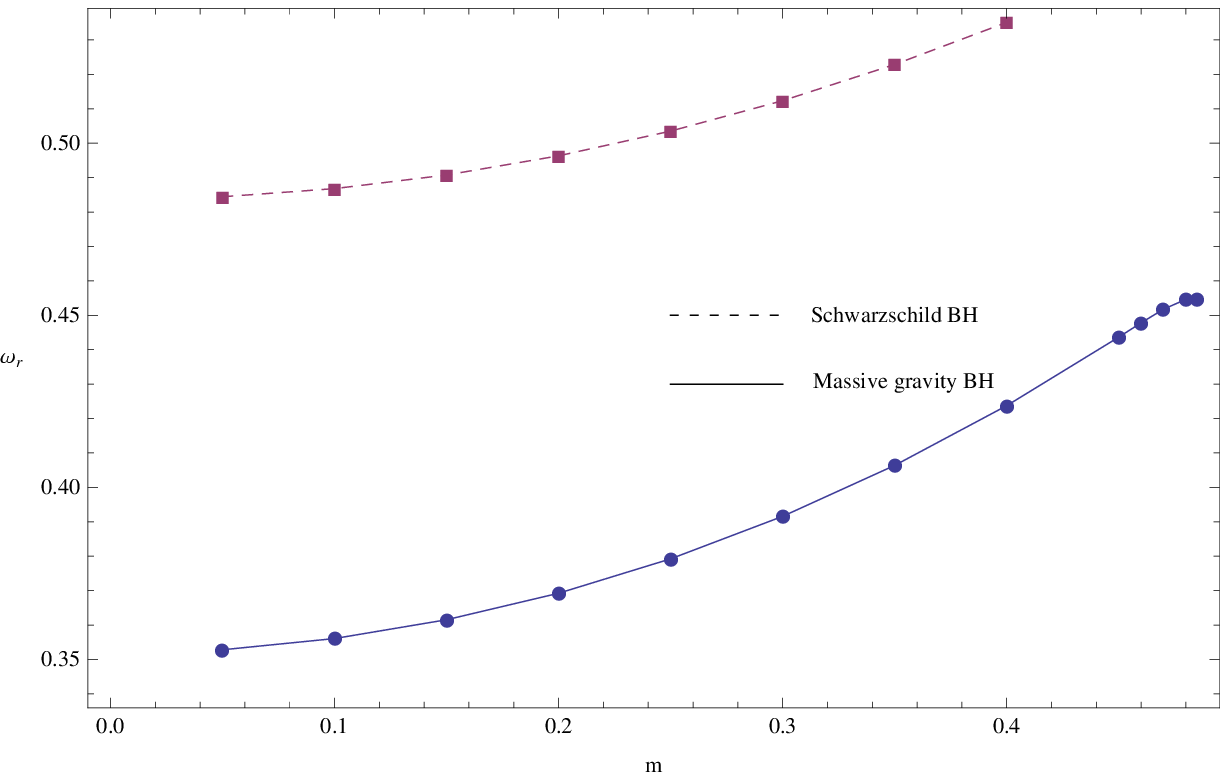}}
\vspace{0.3cm}
 \end{center}

Figure 26. The figure shows the $\omega_R$ vs $m$. Here $M=1, Q=0.75, l = 2$ and $\lambda =1$.\\

The imaginary part of  $\omega$ for the massive scalar field is plotted in Fig. 27 for both black holes. $\omega_I$ is greater for the Schwarzschild black hole compared to the massive gravity black hole. Hence the field will decay faster around the Schwarzschild black hole. Also, the massless field has higher $\omega_I$ than the ones with the mass. Hence the massive  field decays slower. Interestingly, when the mass is increased, there is a critical point where $\omega_I$ approaches zero leading to purely real modes. Hence, quasiresonance modes exists for the massive gravity black hole also.

\begin{center}
\scalebox{.9}{\includegraphics{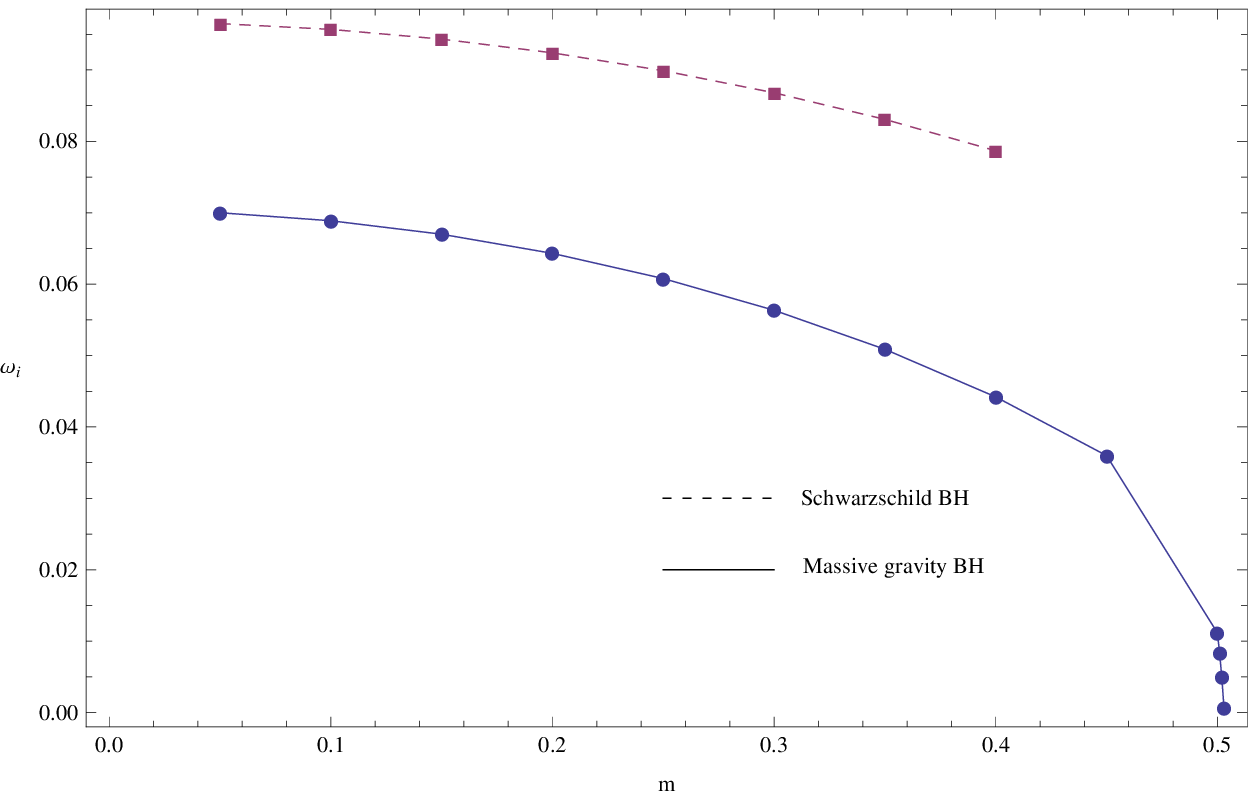}}
\vspace{0.3cm}
 \end{center}

Figure 27. The figure shows the $\omega_I$ vs $m$. Here $M=1, Q=0.75, l = 2$ and $\lambda =1$.\\


\section{Conclusion}

We have studied  scalar perturbations of  a black hole in massive  gravity. First, we studied  QNM frequencies of the massless scalar field via the WKB approach by varying various parameters in the theory.  For small values of $\lambda$  both $\omega_R$ and $\omega_I$ increases. When $\lambda$ gets larger, $\omega_R$ reaches a stable value. On the other hand, when $\lambda$ increases, $\omega_I$ reaches a maximum value and then decreases to a stable value. This stable value is the $\omega$ corresponding to the Schwarzschild black hole with the same mass (with same $l$ value). Hence, the massive gravity black hole decays slower than the Schwarzschild black hole except for a small range of $\lambda$ values. 

We used P$\ddot{o}$schl-Teller approximation to find an analytical approximation for the $\omega$ when $\lambda$ is large. Our computation was done for large $l$.  Both $\omega_I$ and $\omega_R$ increased to converge to a stable value which was the $\omega$ for the Schwarzschild black hole. This is similar to the behavior for $\omega$ obtained using the WKB numerical approach.

When the spherical harmonics index is increased, $\omega_R$ increases in a linear fashion. We have computed $\omega$ for the second mode ($n=1$) also and observed similar behavior.  $\omega_I$ on the other hand decreases with $l$ and approximate a constant value for large $l$. Again for $n=1$, similar behavior persists.

In a separate section we have also computed QNM frequencies via the unstable null geodesics approach.  $\omega_R$ behaves similar to the behavior predicted by the WKB approach. However, $\omega_I$ has similar approach as the WKB approach only for the initial $Q$ values. After a certain range, $\omega_I$ decreases which was not observed by the WKB approach. .

Finally, we computed QNM frequencies of the massive scalar field.  When the mass of the scalar field $m$ is increased, $\omega_R$ is increased. $\omega_R$ for the black hole in massive gravity is smaller than for the Schwarzschild black hole.  When $m$ is increased, $\omega_I$ decreases and it is smaller than the values for the Schwarzschild black hole. Also, when the mass $m$ increased, there is a point where $\omega_I$ reaches zero leading to a state of {\it quasiresonance}. Similar behavior is exhibited by the Schwarzschild black hole for the massive scalar field.

Generally speaking, for all parameters, $\omega_I$ was negative. That means the black hole was stable under massless as well as massive scalar field perturbations.

In continuing this work, It would be interesting to study the behavior of the QNM frequencies for $Q<0$ values and compare with the Schwarzschild black hole. Dirac field perturbation is also would be an interesting avenue to proceed.


\end{document}